\documentclass[sigconf, nonacm]{acmart}

\usepackage{multirow}
\usepackage{fancyhdr}

\fancypagestyle{firstpage}{%
  \fancyhf{} 
  \fancyhead[L]{\textbf{\textcolor{red}{PREPRINT}}} 
}

\setcopyright{acmlicensed}
\copyrightyear{2018}
\acmYear{2018}
\acmDOI{XXXXXXX.XXXXXXX}
\acmConference[Conference acronym 'XX]{}{}

\begin{document}

\title{DiminishAR: Diminishing Visual Distractions via Holographic AR Displays}

\author{JangHyeon Lee}
\email{lee04588@umn.edu}
\orcid{0000-0003-4447-3697}
\affiliation{%
  \institution{University of Minnesota}
  \city{Minneapolis}
  \state{MN}
  \country{USA}
}
\affiliation{%
  \institution{Simon Fraser University}
  \city{Burnaby}
  \state{BC}
  \country{Canada}
}
\author{Lawrence H Kim}
\email{lawkim@sfu.ca}
\orcid{0000-0003-1278-5688}
\affiliation{%
  \institution{Simon Fraser University}
  \city{Burnaby}
  \state{BC}
  \country{Canada}
}

\renewcommand{\shortauthors}{Lee and Kim}

\begin{abstract}
Smartphones are integral to modern life, yet research highlights the cognitive drawbacks associated with their mere presence. While physically removing them can mitigate these effects, it is often inconvenient and may heighten anxiety due to prolonged separation. To address this, we use holographic augmented reality (AR) displays to visually diminish distractions with two interventions: 1) Visual Camouflage, which disguises the smartphone with a hologram that matches its size and blends with the background, making it less noticeable, and 2) Visual Substitution, which occludes the smartphone with a contextually relevant hologram, like books on a desk. In a study with 60 participants, we compared cognitive performance with the smartphone nearby, remote, and visually diminished by our AR interventions. Our findings show that the interventions significantly reduce cognitive impairment, with effects comparable to physically removing the smartphone. The adaptability of our approach opens new avenues to manage visual distractions in daily life.
\end{abstract}


\begin{teaserfigure}
    \includegraphics[width=\textwidth]{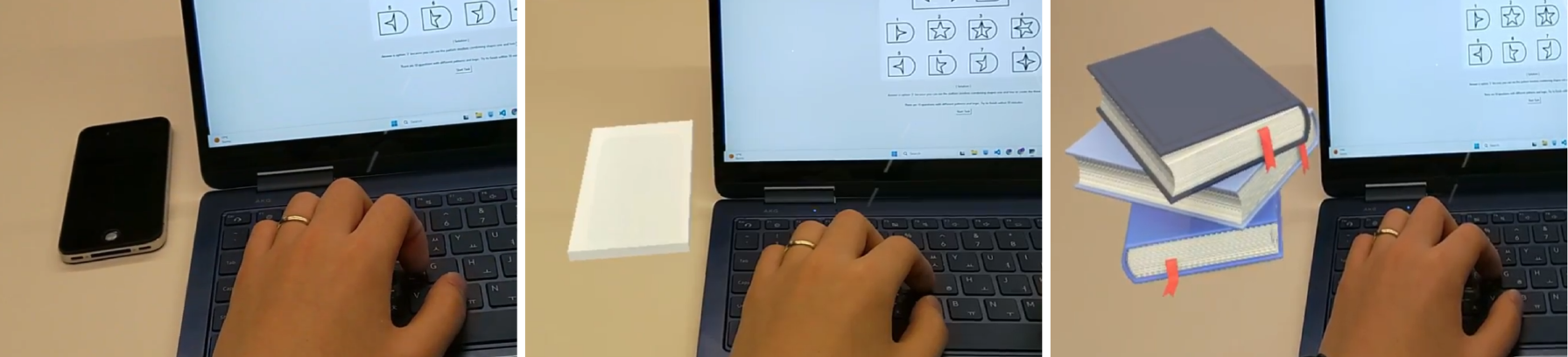}
    \caption{AR serves as a visual (noise) canceling technique to mitigate cognitive impairment caused by distractions (e.g., smartphones). \textbf{Visual Camouflage (center)} disguises the features of the smartphone through a projection of a hologram that closely matches the object's dimensions and mimics the background's color and texture, making the smartphone less noticeable. \textbf{Visual Substitution (right)} modifies the immediate view of the smartphone with a contextually congruent hologram, such as books in a desk environment, subtly altering the user's perception of the scene.}
    \Description{A person is working on a laptop with a phone next to it. The middle image shows a cuboid hologram projected on a phone, camouflaging the phone underneath. The right image displays a stack of books projected on the phone, replacing the immediate view of the phone with a contextually appropriate virtual object.}
    \label{fig:vis}
\end{teaserfigure}
\received{20 February 2007}
\received[revised]{12 March 2009}
\received[accepted]{5 June 2009}
\maketitle
\thispagestyle{firstpage}  
\section{Introduction}
The ubiquity of smartphones offers constant connectivity but at a cost. This cost, known as the \textit{brain drain} effect \cite{ward2017brain}, depletes mental resources and impairs cognitive functions, including memory, attention, and executive processing \cite{kool2010decision}. This depletion directly impacts cognitive performance, weakening the ability to process information, make decisions, and maintain focus on tasks \cite{lane1982limited}.


Research consistently shows that even the mere presence of a smartphone can diminish task performance \cite{thornton2014mere, ward2017brain, johannes2018hard, tanil2020mobile, skowronek2023mere}. This raises a question: \textit{If the presence of a smartphone reduces cognitive performance, could visually diminishing its presence via AR mitigate this cognitive impairment?}

AR has emerged as a transformative technology with significant implications for managing cognitive load, aligning well with Cognitive Load Theory (CLT) \cite{sweller2011cognitive}. CLT explains how the human brain processes information and posits that short-term memory has a limited capacity, capable of processing only a certain amount of information at once. By strategically presenting information and enhancing user environments, AR has shown the potential to reduce cognitive load, improving learning, decision-making, and user experience across various domains \cite{jeffri2021review, buchner2021systematic, buchner2022impact}. This versatility is evident in applications ranging from education and therapy to consumer behavior, thereby enriching cognitive well-being and user experiences \cite{siu2018investigating, desai2016augmented, yoon2017augmented, zhang2019using, keller2021cognitive, barta2023using}.

In contrast to AR's traditional role in reducing cognitive load by amplifying the environment to support user experience, our work uses the concept of diminished reality (DR) to reduce the visual salience of distracting elements, thereby directly targeting and minimizing extraneous cognitive load. This visual cancellation can be particularly effective in environments where excess visual information could lead to cognitive overload, distraction, or decreased efficiency in task completion. Using the Microsoft HoloLens 2 (HL2), we implement two interventions for reducing the visual salience of a smartphone: visual camouflage and visual substitution (Fig. \ref{fig:vis}). The visual camouflage technique projects a customized cuboid hologram over a smartphone to blend with the background, hiding it from the user's field of view (FOV). In contrast, visual substitution covers the smartphone with a context-appropriate hologram in harmony with the environment. These techniques aim to reduce visual distractions, much like noise-canceling headphones that eliminate auditory distractions.

By visually camouflaging or substituting distractions (e.g., smartphones) from the user's FOV using AR holograms, we aim to mitigate the \textit{brain drain} effect \cite{ward2017brain}. This approach is grounded in the principle that visual clutter impairs perceptual clarity and increases judgment errors, emphasizing the importance of strategies that enhance focus in cluttered environments \cite{baldassi2006visual}. Moreover, empirical investigations into the direct cognitive benefits of DR, particularly in the context of mitigating smartphone-induced distractions, remain unexplored \cite{murph2021diminishing, murph2022methods}. This gap motivates our research questions: Do the holographic AR interventions (Fig. \ref{fig:vis}) lead to better cognitive performance than a physically nearby phone, and do they achieve similar performance levels as when the phone is physically removed? 

To address these questions, we investigate the cognitive effects of visual camouflage and substitution through a series of standardized tasks: Operation Span (OSPAN) \cite{unsworth2005automated}, Raven's Standard Progressive Matrices (RSPM) \cite{raven1998court}, and the Go/No-Go (GNG) \cite{bezdjian2009assessing}. These tasks are selected based on their proven effectiveness in measuring cognitive capacity and sustained attention \cite{ward2017brain}. We find that our AR interventions significantly improve cognitive performance to levels comparable to those when the smartphone is physically removed.

Using the HL2, our work blends AR's augmentation capabilities with DR's focus on reducing visual clutter. We present the HCI community with design strategies that cater to streamlining user environments for cognitive benefits. Through several design iterations, we provide detailed guidelines that AR developers can immediately apply to enhance cognitive performance in everyday working environments. While some prior work has explored methods to eliminate distractions \cite{tsurukawa2015filtering, koshi2019augmented, murph2021diminishing, murph2022methods, cheng2022towards, dutt2024kawach}, to the best of our knowledge, this is the first paper to streamline the design space of AR holograms specifically for distraction reduction and to demonstrate significant empirical cognitive benefits from visually canceling ubiquitous devices like smartphones in an optical see-through head-mounted (OST-HMD) AR setting. While our study focuses on smartphones due to their omnipresence, the method is adaptable to any distracting objects.
\section{Related Work}
This section provides an overview of the relevant research in three domains: the cognitive effects of smartphone presence, the use of AR to improve cognitive well-being, and the role of DR on cognitive well-being.
\subsection{Exploring the Cognitive Hazards of Smartphone Presence}
There is consistent research that reveals the adverse cognitive side effects of smartphones. Initial investigations reported the potential distraction of smartphones, noting significant declines in task performance due to their mere presence \cite{thornton2014mere}. Subsequent studies expanded this narrative, demonstrating how the proximity of smartphones could impair working memory and attention, regardless of their power conditions \cite{ward2017brain, skowronek2023mere}. The concept of smartphone vigilance, where the visibility of smartphone notifications hinders our ability to focus on other tasks, further explains how visible smartphones compromise our concentration \cite{johannes2018hard}. Moreover, connections were made between intensive smartphone usage and broader issues such as declines in academic performance, self-control deficits, and adverse mental health outcomes, including depression and anxiety \cite{kim2019understanding, sunday2021effects, wacks2021excessive, daniyal2022relationship, fabio2022problematic}. Excessive smartphone engagement has been linked to a higher incidence of cognitive failures, highlighting the devices' capacity to monopolize cognitive resources and degrade cognitive performance \cite{hartanto2023smartphone, skowronek2023mere}. This body of evidence collectively paints a concerning picture of the cognitive hazards posed by smartphones, emphasizing the importance of addressing this issue in our digitalized society \cite{tanil2020mobile}.

In response, our study introduces a targeted approach to alleviate the cognitive costs associated with the presence of smartphones. We propose using AR to visually diminish the presence of smartphones, showcasing how AR can be a practical tool to mitigate daily distractions.
\subsection{Augmented Reality (AR) for Cognitive Well-being}
AR plays a significant role in cognitive load management, closely aligning with Cognitive Load Theory (CLT), which categorizes cognitive load into intrinsic, extraneous, and germane \cite{sweller2011cognitive}. Intrinsic load refers to the inherent complexity of the task at hand, the extraneous load is associated with how information is presented, and germane load involves the mental effort required to integrate new knowledge into existing frameworks. AR enhances cognitive performance by amplifying the real-world environment with additional virtual objects, which can reduce extraneous load and support germane load by providing contextually relevant information that aids in education \cite{yoon2017augmented, zhang2019using, thees2020effects, keller2021cognitive, uriarte2022higher}. Beyond educational contexts, AR reduces cognitive dissonance, enhancing purchase intentions in consumer behavior \cite{barta2023using}, and extends to health and well-being by supporting psychiatric training \cite{chiam2021novel}, stroke rehabilitation \cite{desai2016augmented}, and driving safety for the elderly \cite{schall2013augmented}. Even popular AR games like Pokémon GO have demonstrated cognitive and social benefits \cite{ruiz2018effect}, while AR pets provide a form of companionship for older adults \cite{cho2021study}. These findings underscore AR's potential to enhance learning, user experience, and well-being through contextually relevant virtual overlays.

In contrast to AR's role in reducing cognitive load by amplifying the environment to support user experience, our work uses diminished reality techniques to reduce distracting elements, thereby directly reducing extraneous cognitive load. This approach is especially beneficial in settings where excessive visual information might cause cognitive overload or hinder task efficiency.
\subsection{Diminished Reality (DR) for Cognitive Well-being}
DR, like AR, involves visual manipulation of the world, but its focus is fundamentally different. While AR overlays additional virtual elements to enhance interaction and creativity, DR concentrates on diminishing or removing specific elements to simplify perception and reduce distractions, emphasizing the minimization of visual saliency rather than augmentation \cite{Steve1999, herling2010advanced, mori2017survey}. By visually removing or occluding non-essential elements, DR helps users focus on the most relevant aspects of their environment \cite{cheng2022towards}. This approach offers distinct advantages over physically removing omnipresent distractions like smartphones. By diminishing the device’s visual salience, DR reduces extraneous cognitive load (the mental effort spent on irrelevant information) while still allowing user access. This balance maintains situational control, avoiding the anxiety or inconvenience associated with prolonged complete removal \cite{hartanto2016smartphone}. Several studies have also examined DR's role in stress and workload management \cite{murph2021diminishing, chan2022declutterar}, skill training \cite{sakai2018d, murph2022methods}, product design \cite{siltanen2017diminished, pfaff2023reality}, privacy \cite{tabet2023mobile}, interaction quality \cite{yao2013focalspace, vuarnesson2021shared}, and user experience in hand-held AR settings \cite{kim2020don, kim2023real}. 

Despite these advances, current research lacks empirical evidence on how DR improves cognitive performance by visually eliminating distractions. Unlike prior work that addresses broad applications, our study aims to mitigate the cognitive decline linked to smartphone presence and uses DR in a holographic AR setup to either visually camouflage or substitute smartphones.
\section{System Design \& Implementation}
We illustrate our approach, focusing on how we achieve visual cancellation of distractions and the design considerations for delivering optimal AR experience. We also narrate our investigation through various methods before arriving at the most effective approach for our study. This process was essential in shaping the final design and execution of our experiment, helping us identify the potential and constraints associated with each method explored.
\subsection{Visual (Noise) Cancellation}
We aim to address two design goals. First, we want to \textit{cancel out} visual distractions like how noise-canceling headphones reduce auditory distractions. Second, we seek to develop AR holograms that integrate into the environment without becoming distractions themselves. This leads to two techniques: Visual Camouflage and Visual Substitution (Fig. \ref{fig:vis}). Although our study focuses on diminishing the visual salience of smartphones, this method is generalizable and can be applied to other distracting objects as well.
\subsubsection{Visual Camouflage}
Similar to how noise cancellation creates an anti-noise wave, visual camouflage uses a hologram that matches the shape and size of the distracting object but alters its visual features, such as color and texture, to blend with the background, reducing the object's visual salience. Specifically, we project a cuboid hologram, customized to replicate the background's visual features, slightly larger than the smartphone to ensure complete coverage. We capture an image of the empty workspace and extract its features to achieve this. When the smartphone is placed in the workspace, we overlay it with a customized hologram that matches the pre-captured background.
\subsubsection{Visual Substitution}
Rather than camouflaging the smartphone to blend with the background, visual substitution occludes the object of interest (i.e., smartphone) with a contextually appropriate hologram that harmonizes with the environment, such as a book on a desk. Here, we not only block the immediate view of the smartphone but also introduce study-related objects, like books, creating an environment that potentially promotes focus by repurposing the distraction. Therefore, while visual camouflage creates a distraction-free zone, visual substitution modifies the distraction and repurposes the space to enhance focus or task relevance.
\subsection{AR Systems vs. VR Systems}\label{sec:ar}
To closely evaluate cognitive performance in environments mirroring real-world scenarios, we designed our experimental setup to emulate typical desk-based tasks. Such an approach demanded real-time technology that could blend digital elements with the physical world, retaining an authentic connection to the user's immediate environment. AR displays emerged as the prime choice, influenced by several technical and experiential factors over VR settings.

Optical see-through head-mounted displays (OST-HMDs) or AR HMDs allow virtual elements (i.e., holograms) to be overlaid directly onto a user's view of the real world. On the other hand, video passthrough (VPT) HMDs or VR HMDs recreate a user's environment within the virtual space. Though this VPT approach addresses challenges related to occlusion and the limited FOV of OST-HMDs, they are not devoid of limitations. These VR HMDs have been associated with introducing real-scene distortions and unstable visual experiences \cite{anthes2016state, stauffert2020latency}. Additionally, their resolution often lacks the clarity and detail of the real world, as they project surroundings onto a pixelated screen, creating a disconnect from reality. VR HMDs often suffer from system latency, causing temporal inconsistencies between the user's actions and the system's responses \cite{baumeister2017cognitive}. This mismatch can disrupt cognitive tasks and affect the validity of experiments.

In contrast, AR's direct see-through of the environment minimizes latency issues since it avoids the need for a virtual rendering of the real world \cite{nabiyouni2017relative}. This provides a more consistent user experience, enabling an accurate assessment of cognitive functions by resembling normal glasses that offer a clear view of the real environment \cite{itoh2021towards}. While VR's immersive capabilities are promising, its inherent trade-offs made AR systems more fitting for our study. AR's ability to combine virtual interventions with the real world without significant discrepancies ensures that participants' cognitive functions are assessed with less confounding variables.
\subsection{Hologram Design Space}
The effectiveness of our AR interventions is closely tied to the technical capabilities of the Microsoft HoloLens 2 (HL2), an optical see-through device \cite{itoh2021towards}. To achieve an optimal visual cancellation of a phone via AR, careful hologram design is essential. To better understand the design space, we conducted a design exploration with 5 participants, including 2 members of the research team. The primary objective was to understand how to design an AR hologram that minimizes the visual saliency of the phone while considering the constraints of the HL2 device. Participants interacted with the system to evaluate the hologram design, including color, texture, size, dimension, quantity, and animation. These interactions offered valuable insights, which informed our design decisions and are detailed in the following sections.

\begin{figure}[h!]
    \centering
    \includegraphics[width=\linewidth]{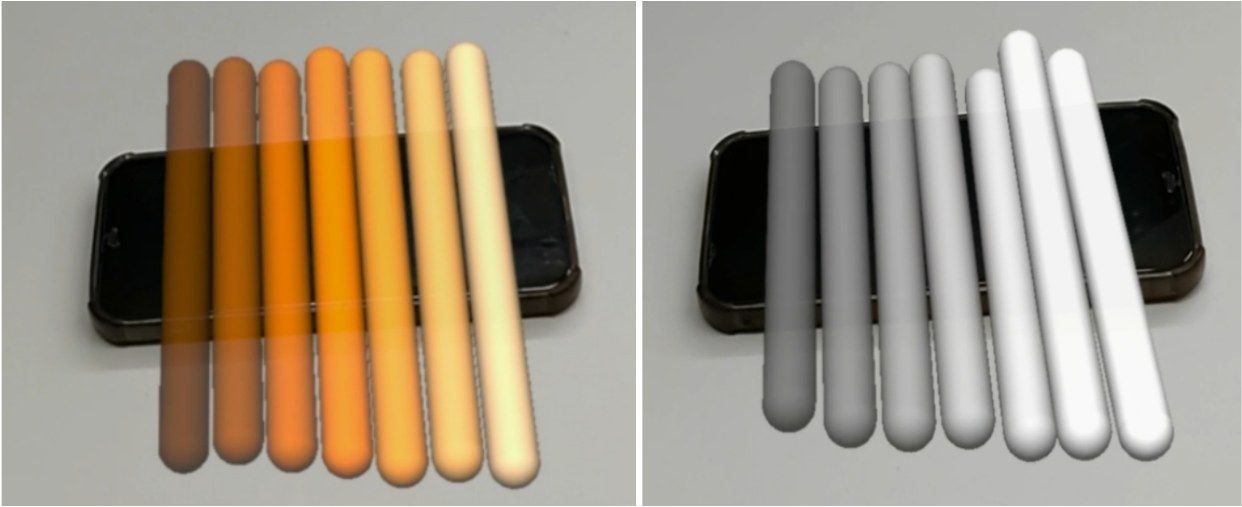}
    \caption{Left image shows a gradient of orange shades from dark to bright demonstrates the impact of color on visibility, with brighter hues providing better concealment. Right image shows a gradient in grayscale, where brighter whites obscure the underlying object compared to darker shades.}
    \Description{The images illustrate how hologram brightness affects the concealment of a smartphone. The left image shows a gradient of holograms in shades from dark orange to bright orange atop a smartphone, demonstrating the variation in concealment effectiveness. The right image shows a gradient from dark grey to bright white holograms, indicating how increased brightness can render underlying objects less transparent and more effectively occluded.}
    \label{fig:color}
\end{figure}

\subsubsection{Color}
In holographic displays, the augmentation of the real world involves adding light, which results in darker or black colors appearing more transparent than brighter or white colors \cite{erickson2020exploring}. To evaluate color renderings on the HL2, we projected a spectrum of holograms on a smartphone: from bright orange to dark brown and from bright white to dark grey (Fig. \ref{fig:color}). Our findings show that darker holograms are less effective for concealment due to their increased transparency on AR displays, leading to a preference for brighter hologram colors for optimal visual cancellation.

\begin{figure}[h!]
    \centering
    \includegraphics[width=\linewidth]{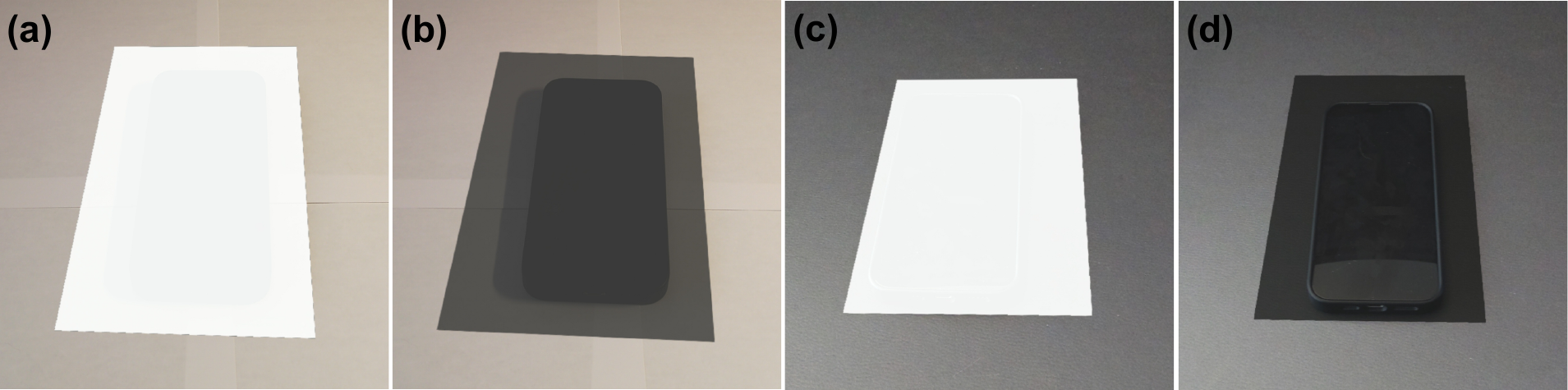}
    \caption{(a) shows a white hologram with a white surface, concealing the phone. (b) shows a dark hologram on a white surface, where the phone is translucent. (c) shows a white hologram on a dark surface, where the hologram becomes prominent. (d) shows a dark hologram on a dark surface, where translucency reveals the phone, demonstrating the limited effectiveness of darker holograms for concealment.}
    \Description{The images compare the effectiveness of hologram colors against different background contrasts. (a) shows a white hologram concealing a phone on a white surface. (b) shows a dark hologram on a white background, where the phone is not as well hidden. In (c), a white hologram on a dark surface makes the phone not visible. (d) shows a dark hologram on a dark surface, where the phone is visible, demonstrating that bright holograms offer the best concealment. These results emphasize the importance of matching hologram color to the background context for optimal visual cancellation.}
    \label{fig:pair}
\end{figure}

We further tested the color rendering against contrasting backgrounds (Fig. \ref{fig:pair}b and Fig. \ref{fig:pair}c). We found that brighter holograms were rendered more efficiently against a darker background. This phenomenon can be attributed to the principle of visual contrast \cite{blaha2014colour}, where the juxtaposition of a bright element against a dark backdrop accentuates the former. However, one would choose a hologram color analogous to the background for optimal blending. Since the HL2's additive display rendered darker colors translucent, we limited our choice to a pair of bright holograms with a bright background. Specifically, we use a simple white desk mat to fine-tune the hologram's color to blend with this white background.

\subsubsection{Texture}
\begin{figure}[h!]
    \centering
    \includegraphics[width=\linewidth]{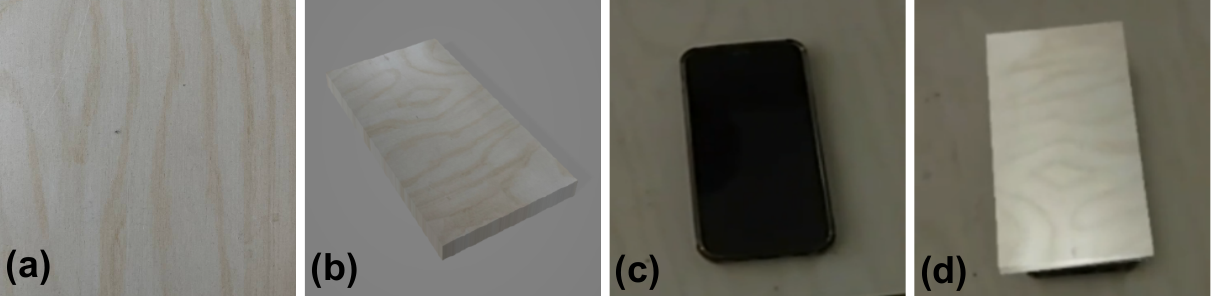}
    \caption{(a) shows a wood grain surface. (b) depicts the cuboid hologram during inpainting process with the wood grain texture. (c) displays a smartphone placed on the wood surface. (d) shows the smartphone covered by a hologram with the wood texture, indicating the importance of texture alignment to avoid the unnatural appearance due to mismatched patterns.}
    \Description{The sequence demonstrates the visual camouflage process on a complex wood grain texture. (a) displays a wood grain surface, establishing the background pattern. (b) shows a 3D cuboid hologram rendered with a wood grain texture. (c) includes a smartphone placed on the wood surface. (d) depicts the smartphone partially camouflaged with a hologram matching the wood grain texture.}
    \label{fig:text}
\end{figure}

The fidelity of a hologram depends on its texture. Shadows, patterns, and detailed nuances are crucial for making a hologram appear realistic \cite{pardo2018correlation}. We use the 3D Builder application to apply 2D images onto 3D models, creating environment-specific textures. However, reflective surfaces and intricate patterns, like wood grain, can reduce a hologram's fidelity if the reconstruction does not match the background accurately (Fig. \ref{fig:text}d). Hence, a simple, non-reflective, plain white background is preferred. This neutral choice simplifies replication and reduces discrepancies between the hologram and its environment, optimizing the overall realism and efficacy of the AR experience.

\subsubsection{Size}
\begin{figure}[h!]
    \centering
    \includegraphics[width=\linewidth]{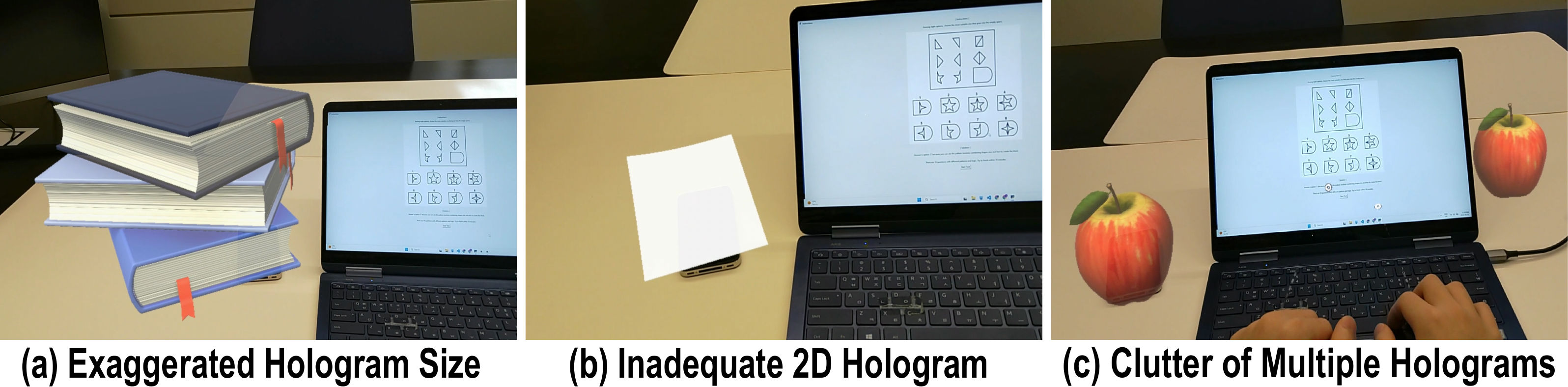}
    \caption{(a) The size of book hologram is exaggerated, disrupting the visual coherence of the scene. (b) Despite the smartphone's thinness, the phone's three-dimensionality is apparent, making the 2D approach ineffective. (c) Multiple holograms diverge focus, opting for a single hologram scene.}
    \Description{(a) shows an AR hologram of three large books stacked on a desk beside a laptop. The books are significantly oversized relative to the surrounding objects, highlighting the importance of maintaining natural proportions in AR applications. (b) shows a flat 2D white hologram attempting to cover a smartphone on a desk next to a laptop. The hologram hovers above the smartphone. (c) shows a laptop on a desk with two holographic apples on either side of the screen.}
    \label{fig:combo}
\end{figure}

Accurate hologram sizing is crucial for effective visual interventions. The hologram's dimensions must mirror or exceed the phone's, as seen from the viewer's perspective. This ensures the phone is entirely obscured from the visual field. For visual substitution, the book hologram must cover the phone and fit naturally within the environment. If the hologram matches the phone's dimensions too closely, it may appear unnatural, as books typically differ in size from smartphones. Conversely, if the book hologram is too large, it disrupts the realism of the scene (Fig. \ref{fig:combo}a). This effect relates to the "Big Baby" effect {\cite{anjos2019adventures}}, where disproportionate scaling distorts the perception of natural size relationships. While the "Big Baby" effect originally describes distortions in human representation, a similar principle applies here. A book hologram that is either too large or too small relative to its surroundings appears unnatural, reducing immersion and perceptual realism. While it is crucial for the hologram to cover the device, its size should not distort the natural environment of the workspace.

\subsubsection{Dimension}
Inadequate dimensionality can also reduce the effectiveness of visual canceling techniques (Fig. \ref{fig:combo}b). Specifically, the human cognitive system could display a level of skepticism towards the holographic overlay if it lacks the depth cues that the human eye is accustomed to \cite{diaz2017designing}. For instance, a 2D hologram, such as a virtual piece of paper, would appear to hover above the smartphone rather than fully conceal it, undermining the intended effect. This observation underscores the importance of prioritizing 3D designs for real-world objects that are typically 3D \cite{li2023comparative}, ensuring that AR holograms integrate seamlessly with the environment and effectively achieve the desired cancellation.

\subsubsection{Quantity}
The number of holographic elements was controlled to include only a single hologram (Fig. \ref{fig:combo}c). This decision aligns with the concept of perceptual load, which refers to the amount of visual information presented and is related to the levels of clutter within a scene \cite{rosenholtz2007measuring}. Prior research on visual clutter suggests that excessive information in a scene can decrease recognition performance \cite{rosenholtz2007measuring}. Therefore, increasing the number of objects in a scene can raise cognitive load. By limiting the scene to one targeted hologram on the smartphone, we aimed to minimize cognitive load and enhance the effectiveness of the visual canceling technique.

\subsubsection{Animation}
\begin{figure}[h!]
    \centering
    \includegraphics[width=\linewidth]{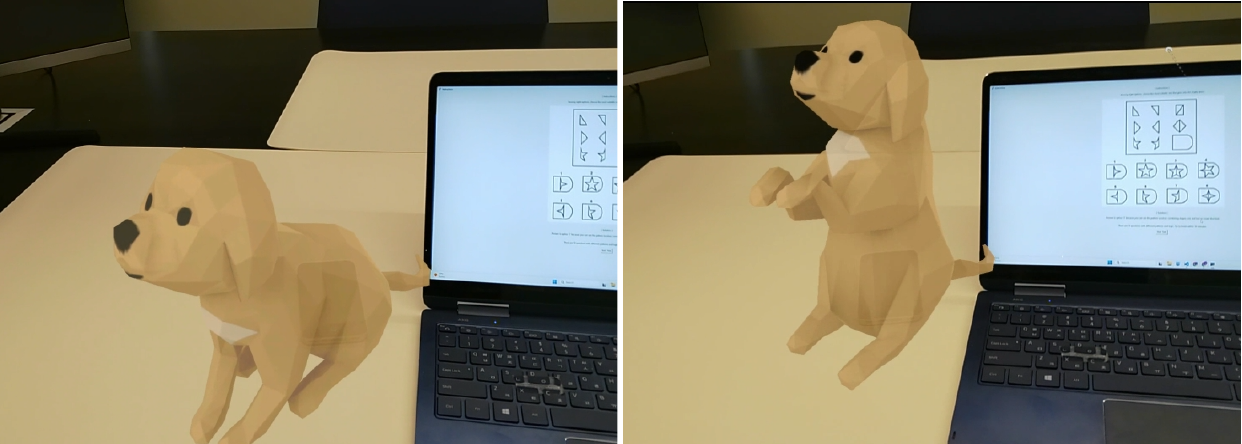}
    \caption{Static dog hologram (left) is still, while dynamic dog hologram (right) constantly barks with movements.}
    \Description{The left image features a static hologram posing no movement, while the right image displays the dynamic hologram engaging in a barking motion. The static hologram was chosen to reduce cognitive load, aiming for a simpler user experience.}
    \label{fig:animia}
\end{figure}

We chose static holograms to minimize the potential cognitive load associated with moving holograms for our study (Fig. \ref{fig:animia}). Previous research has shown that animations often provide no advantages over still images \cite{tversky2002animation}. In fact, according to Mayer’s Cognitive Theory, animations can lead to increased cognitive load by taxing cognitive resources with unnecessary motion that do not directly contribute to the task at hand \cite{mayer2005principles}. Using static holograms, we aim to focus user's attention on the essential elements of the scene without overloading their cognitive capacity, thus enhancing the effectiveness of visual cancellation. This approach ensures that the holograms serve their intended purpose without introducing additional distractions or cognitive strain.

\subsection{Hologram Placement \& Sizing}
In our exploration of automating the hologram placement process on the smartphone, we investigated several methods for real-world object detection. Ensuring the hologram is accurately and automatically placed on the desired object requires real-time performance and robust detection, regardless of the phone type. We discuss four methods we considered, detailing their requirements and drawbacks. Additionally, we address the management of hologram size after placement to ensure optimal integration with the target object.

Microsoft Azure Object Anchors method requires converting a 3D model to an object anchors model of the real-world object generated through the Microsoft Azure Cloud's conversion service. The object anchors model serves as a tracking reference. Using the conversion service, we aimed to detect real-world objects with the object detection SDK in HL2. However, the conversion service only supports more substantial objects, ranging from 1m to 10m, becoming ineffective for smaller items like smartphones.

VisionLib's Object Tracking method requires a 3D model of the target object. Although the 3D model appeared floating upon initiating the application, directing one's gaze toward the actual phone anchored the model to it. While successfully detecting and overlaying holograms onto the smartphone, this method presented usability challenges. Users had to expend effort to lock the 3D model onto their smartphone. The phone's slim dimension made it challenging to detect due to its lack of distinct features, akin to a 2D object. Moreover, 2D tracking struggled because the phone screen is reflective and lacks distinctive features for reliable detection.

Vuforia Object Recognition demanded detailed prior knowledge of the object's dimensions and necessitated an Android device for scanning via Vuforia's app. Again, the thin dimensions of most phones made it difficult to pinpoint the correct aspect ratio without inducing system errors. Attempting to accommodate these variable dimensions during experiments could lead to unintended delays, potentially skewing the user experience.

Lastly, while ArUco markers are widely used for their robustness as placeholders \cite{hubner2018marker}, their application within our study was deemed intrusive. In our study, placing an ArUco marker on a desk or attaching it to a user's phone presents certain disadvantages. Firstly, putting a QR code on a workspace or personal device is uncommon, detracting from the natural environment we aim to replicate. Moreover, the repetitive task of attaching and detaching markers on different users' phones can interrupt the flow of the experiment. Relying on the HL2's capabilities, without external aids (e.g., ArUco), was essential to maintaining a natural setting.

We used a hands-on approach to hologram management (i.e., manual placement and sizing) to provide the most robust AR experience. Given the varying sizes of smartphones (e.g., brand, model, and phone cases), a real-time adjustment of the hologram's size was also essential. We first import the hologram from a 3D library in HL2, followed by a grab gesture to position it over the participant's phone. The hologram was placed to cover the phone completely, eliminating any hovering effect. After the placement, we used a pinch gesture to match the participant's phone dimensions. This led to the most realistic user experience by allowing control over the hologram's placement and dimensions without requiring extraneous tools \cite{hubner2018marker}.
\section{Methods}
The study was approved by the University’s Institutional Review Board, and all participants provided informed consent. Additionally, the study was pre-registered on the Open Science Framework (OSF)\footnote{\href{https://osf.io/7gx64}{https://osf.io/7gx64}}, where all data is openly available\footnote{\href{https://osf.io/8fy43/}{https://osf.io/8fy43/}}.
\subsection{Participants} 
60 participants (aged 20-35 years; 11 female, 49 male) were recruited for our in-person experiment. The majority were Computer Science majors (46), with others from Engineering (5), Business (3), and other academic disciplines (6). Most were East Asian (41), with diverse ethnic backgrounds including Indian (9), Punjabi (2), Central Asian (1), South Asian (1), Hispanic or Latino (1), Middle Eastern or North African (2), Whites (2), and one who chose not to specify their ethnicity. Regarding AR/VR experience, 26 were first-time users, 26 had limited experience, 7 had moderate experience, and 1 participant had extensive experience. The sample size was guided by CHI local standards and relevant prior studies \cite{caine2016local, koshi2019augmented, cheng2022towards}. To be eligible, participants were required to have normal or corrected-to-normal vision and be free of neurological disorders (e.g., migraine,  chronic fatigue). Each participant received compensation of \$15.

\subsection{Experimental Setup}
\begin{figure}[ht]
    \centering
    \includegraphics[width=\linewidth]{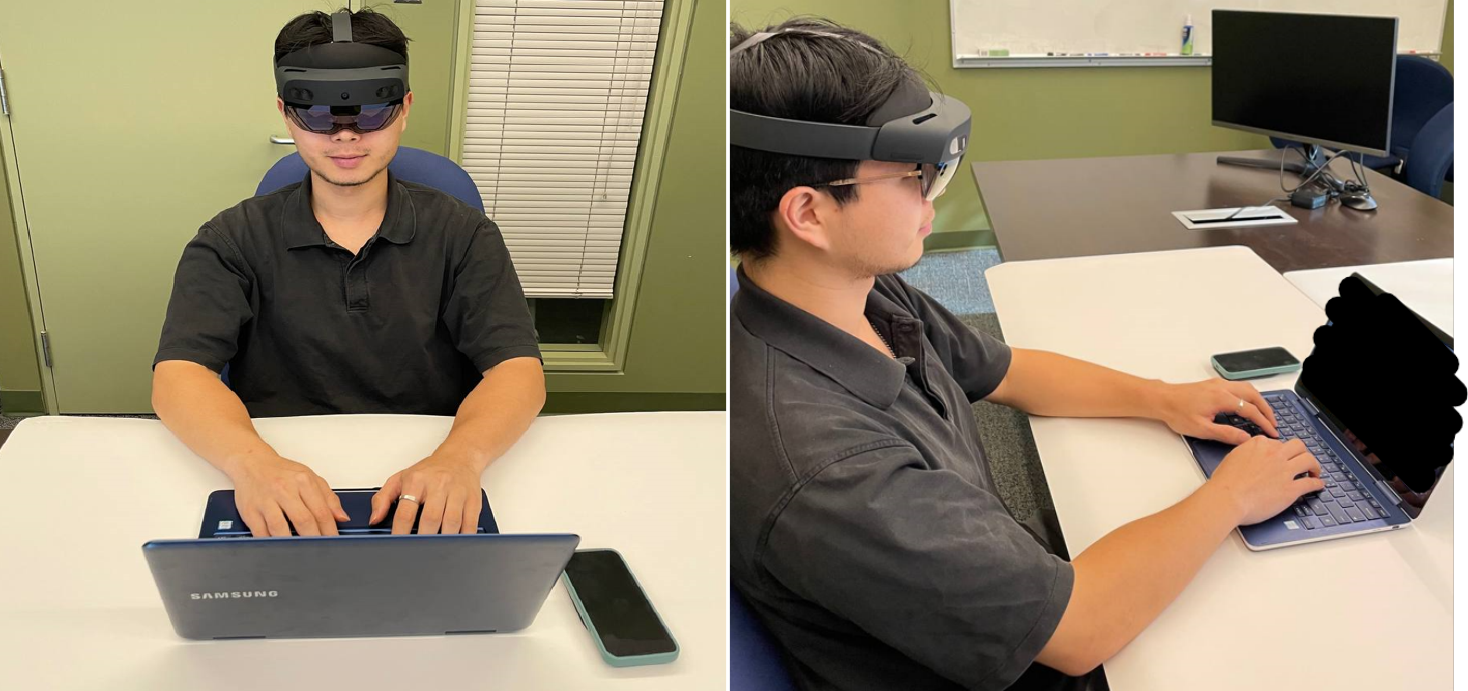}
    \caption{Participants sit with a laptop and smartphone positioned to account for the HL2's limited FOV. This setup ensures the phone stays visible to the participant and the AR system, allowing consistent holographic coverage.}
    \Description{A participant is shown seated at a desk, wearing the AR headset, and interacting with a laptop. The smartphone is positioned near the laptop within the AR device's limited FOV, ensuring it remains visible to both the participant and the AR system for consistent holographic coverage.}
    \label{fig:placement}
\end{figure}

To ensure a consistent user experience during laptop tasks, we provide guidelines to emulate a natural environment. The room's lighting was set to the lowest of three available brightness levels to enhance hologram fidelity by reducing ambient light interference. We installed curtains to prevent external light intrusion, ensuring a consistent visual environment and avoiding potential disruptions in the AR's occlusion capability. A white desk mat was placed on the table to serve as a uniform background for the holograms, particularly for the aspect of visual camouflage (Fig. \ref{fig:placement}).

Given the limited FOV of HL2 (i.e., 52 degrees), precise positioning of elements was critical \cite{bourk2007effects}. When the participant is seated, the smartphone must be close to the laptop and within the user's immediate FOV (Fig. \ref{fig:placement}). This ensures the visual cancellation remains consistent and fully covers the smartphone, even during minor head movements. During the pre-check phase prior to the main study, we observed that extensive seat movement by participants could cause the hologram to fall outside their field of view. This precheck, which mirrored the procedure of the main study, served as an additional quality assurance step to identify and address potential issues for robustness. Addressing this was essential, as it would otherwise defeat the purpose of occlusion by making the smartphone visible. To counteract this, the seating was pre-arranged. A stationary chair, devoid of wheels, was chosen to minimize inadvertent shifts. The chair was equipped with back support to encourage a stable posture, reducing the likelihood of significant head movements.

To ensure consistency across participants and maintain control over the experimental conditions, we instructed all participants to turn their phones off during the experiment. This decision was guided by prior research \cite{ward2017brain, skowronek2023mere}, which found that phone power condition (on vs. off) does not influence cognitive performance. Turning off the phones further minimized unexpected external confounders, such as notification variability from individual usage, app settings, and device behaviors, ensuring our focus remained solely on the effects of minimizing visual salience. While an alternative approach is to provide experimenter-controlled phones to manage notifications, we prioritized ecological validity by allowing participants to use their own devices. This decision reflects the reality that participants are more connected to and familiar with their phones compared to unfamiliar, experimenter-provided devices \cite{konok2017mobile}.
\subsection{Conditions}
Our study uses a between-subjects design with participants randomly assigned to the following conditions:
\begin{itemize}
    \item \textbf{C1} Physically Nearby, where the phone is stationed on the desk.
    \item \textbf{C2} Physically Removed, where the phone is relocated outside the room.
    \item \textbf{C3} Visually Camouflaged, where the phone is disguised with a 3D hologram of the same background features.
    \item \textbf{C4} Visually Substituted, where the phone is occluded with a hologram representing a stack of books.
\end{itemize}

The selection of these conditions was informed by existing research and aims to deepen our understanding of cognitive impacts related to smartphone presence. Specifically, conditions C1 and C2 derive from the \textit{brain drain} study, demonstrating various scenarios of smartphone proximity can significantly influence cognitive capacity \cite{ward2017brain}. We chose the `Physically Removed' condition over alternatives like placing the phone in a bag as this condition showed the largest effect from prior work \cite{ward2017brain}. Conditions C3 and C4 use DR-influenced interventions, motivated by research indicating the potential for transparency and context relevance to reduce distractions, albeit without significant empirical evidence in holographic AR display settings \cite{cheng2022towards}. Prior work suggests that visual changes to the environment may improve concentration and subjective evaluations \cite{cheng2022towards}. From these insights, we hypothesize that the visually camouflaged and substituted phone via AR will enhance cognitive performance compared to a physically nearby phone and achieve similar performance to a physically removed phone.

To address concerns of anxiety from smartphone separation when physically removed, as smartphones are often perceived as an extension of oneself \cite{terzimehic2023tale}, our study was designed to minimize such anxiety. During the recruitment, participants were informed about the session's length and their option to withdraw, reducing potential stress. Literature supports that awareness of separation duration and control over the situation can reduce stress \cite{tams2018smartphone}. Additionally, studies confirm that temporary separation from smartphones does not intensify anxiety or harm well-being, suggesting minimal anxiety impact from phone removal in our study \cite{wilcockson2019digital, brailovskaia2023finding}.
\subsection{Tasks}
The Operation Span (OSPAN) task, which involves math problems and memory sequences, and the Raven's Standard Progressive Matrices (RSPM) task, which focuses on pattern completion, assess cognitive capacity. The Go/No-Go (GNG) task evaluates sustained attention through response to visual cues, differentiating from cognitive capacity measures.

\subsubsection{Operation Span (OSPAN)} The OSPAN task evaluates an individual's capability to retain information in working memory while processing additional unrelated details \cite{engle2002working}. Participants are initially presented with a simple math problem. After solving, participants press either the "C" key if the equation is correct or the "I" key if the equation is incorrect. Immediately following, a random letter is presented, which participants must remember. Each math problem paired with a letter presentation forms a single trial. Trial sets can contain between 3 and 7 trials. After each set, participants recall the letters in the correct sequence. The focus is on both speed and accuracy. In our study, participants undertook five random trials: one for each trial set length (3, 4, 5, 6, and 7). The OSPAN Score, with a maximum of 25, indicates an individual's domain-general attentional resources. It measures the participant's ability to process and store information simultaneously, thereby revealing aspects of working memory. Participants with an accuracy below 85\% on the math operations are excluded \cite{unsworth2005automated}.

\subsubsection{Raven’s Standard Progressive Matrices (RSPM)} The RSPM task measures a non-verbal component of general fluid intelligence that characterizes an individual's capacity to reason and tackle unfamiliar problems \cite{raven1998court}. Participants are shown incomplete pattern matrices and must determine the piece that completes the pattern. Grouped in five 12-item sets (A-E) of escalating difficulty, participants solve ten items: D2, D4, D6, D8, D10, D12, E1, E2, E4, and E6. The RSPM Score, with a max score of 10, is sensitive to the immediate availability of attentional resources due to the task's escalating difficulty. Thus, a high RSPM Score indicates robust attentional control.

\subsubsection{Go/No-Go (GNG)} The GNG task measures sustained attention \cite{o2009two}. Participants respond to sequential "go" and "no go" targets on a computer screen in this task. They press the spacebar for "go" targets and abstain for "no go" targets. Each trial starts with an 800ms fixation point and a 500ms blank screen. A color-changing rectangle cue follows. Participants press the space bar for the green "go" cues and ignore blue "no go" cues, with each cue lasting up to 1000ms. A 700ms gap separates the 100 trials, which are equally divided between "go" and "no go" targets. Metrics are omission errors and reaction time that measure sustained attention without the interference of working memory capacity \cite{redick2011working}. Omission errors track missed "go" responses, serving as a measure of sustained attention. Reaction time measures the speed of responses to "go" targets, indicating attention agility. To handle commission errors, when participants incorrectly respond to "no go" cues, we calculate the commission error rate by dividing the number of commission errors by the total "no go" cues. Participants with rates outside 95\% confidence interval are excluded \cite{littman2017all}.
\subsection{Post-Study Interview}
After the tasks, participants completed a post-study interview using the laptop in front of them. The experimenter remained outside the room to avoid any influence or bias on participant responses. They were asked to rate the visual salience of the phone in the presence of the hologram on a 7-point Likert scale \textit{(Q1: "How visually salient was the phone when the hologram was present?")}. The distinctness of the hologram was assessed similarly \textit{(Q2: "How noticeable was the hologram?")}. Subsequent questions asked the frequency of the participants' attentional shifts toward the hologram \textit{(Q3: "How often did your attention shift to the hologram?")}. Participant inclination toward future adoption of the interventions in related scenarios was then measured, reflecting the hologram's practicality and user acceptance \textit{(Q4: "How likely are you to use the holograms in similar settings in the future?")}. Finally, an open-ended question was given about potential changes in the hologram design \textit{ (Q5: "If you could change the hologram to a different object, what would you change to and why?")}.
\subsection{Procedure}
Each study session spanned an hour on average. Upon entering the lab, conditions were assigned at random to each participant. In addition to prior work reporting that phone power condition does not affect cognitive performance \cite{ward2017brain, skowronek2023mere}, and given the variety of smartphone devices and settings among participants, we asked participants to turn off their smartphones to eliminate any potential confounders beyond the mere presence of their phones. For the physically removed condition, we asked participants to leave all their belongings, including their smartphones, outside the room. Then, participants signed a consent form and completed a pre-study questionnaire. The questionnaire asked for general demographic information and prior experience with AR devices. Simultaneously, the examiner prepared the AR device and holograms for the experiment.

To ensure consistency, all participants wore the HL2, even for participants assigned to conditions without the interventions, thereby eliminating the "sunglass effect" when the device was not worn. The sunglass effect refers to the altered visual perception experienced when not wearing the device, akin to the brightness change when removing sunglasses. Once the participants wore the device, an eye calibration process was initiated to ensure the accurate placement of holograms in the participants' FOV. Then, the participants completed a randomized sequence of the three cognitive tasks to prevent potential order effects. These tasks were conducted in isolation, without the examiner in the same room, to minimize external confounders, such as the examiner's presence \cite{zajonc1965social}. 

After all tasks, participants were interviewed to gather feedback about the interactions with the hologram and the smartphone and their overall experience of the study. Throughout the experiment, the examiner monitored the user’s progress by viewing a live feed of what the user saw through the HL2 device, which was streamed to the examiner's computer via the Windows Device Portal. Google Remote Desktop facilitated the experiment remotely, allowing the examiner to manage the session without entering the participant's room during the tasks.
\subsection{Data Analysis}
We used a multivariate analysis of variance (MANOVA) to evaluate the cognitive capacity effects of different phone conditions on a combination of OSPAN and RSPM scores. To ensure the suitability of parametric tests, we first checked the normality of each cognitive capacity measure (i.e., OSPAN and RSPM) within each condition using the Kolmogorov-Smirnov (KS) test. The results confirmed that the data did not significantly deviate from normality ($p > 0.05$) for all groups. We also verified the homogeneity of variance using Levene’s test for both OSPAN ($p = 0.791$) and RSPM ($p = 0.955$), indicating equal variances across groups. The GNG task was excluded from the MANOVA since it measures sustained attention rather than cognitive capacity. This exclusion aligns with prior work, which shows that sustained attention does not correlate with the cognitive capacities assessed by OSPAN and RSPM \cite{ward2017brain}. For the GNG task, we assessed normality using the same KS test, which revealed that the data did not meet the normality assumption. As a result, we used the non-parametric Kruskal-Wallis test to examine the effects of different phone conditions on two behavioral measures of sustained attention: omission errors and reaction time. Following the MANOVA, we conducted ANOVA tests for each dependent variable (OSPAN and RSPM) to further examine the impact of each phone condition on cognitive capacity. Finally, Bonferroni post-hoc tests explored pairwise differences among the four phone conditions. The p-value was adjusted by dividing the conventional alpha level by the number of pairwise comparisons made, excluding the comparison between conditions C3 (Visually Camouflaged) and C4 (Visually Substituted) as predetermined. All statistical analyses were pre-registered on the OSF, ensuring the integrity of our experiment. No participants were excluded as none met the exclusion criteria.

For the post-study interview, participants responded to four closed-ended questions on a 7-point Likert scale and one open-ended question regarding hologram design preferences. We calculated descriptive statistics, including the mean ($\mu$) and standard deviation ($\sigma$), to summarize participant perceptions of phone visibility, hologram noticeability, attention shifts, and future adoption likelihood. For the open-ended responses on alternative hologram designs, we conducted a brief thematic analysis to identify key patterns and user preferences. One member of the research team conducted the initial review and coding of responses, and a second member cross-checked and confirmed the identified themes. This dual-review process ensured the accuracy and consistency of the analysis. Through this approach, we identified key themes related to functionality, aesthetic appeal, and emotional comfort.
\section{Results}
This section provides a comprehensive view of our study's findings, segregated into task performance and post-interview analysis. Task performance results offer statistical measures of cognitive capacity and sustained attention, while the survey data elucidated the user experience and perceptions through analyses of the post-interview responses.

\begin{figure*}[ht]
    \centering
    \includegraphics[width=0.85\textwidth]{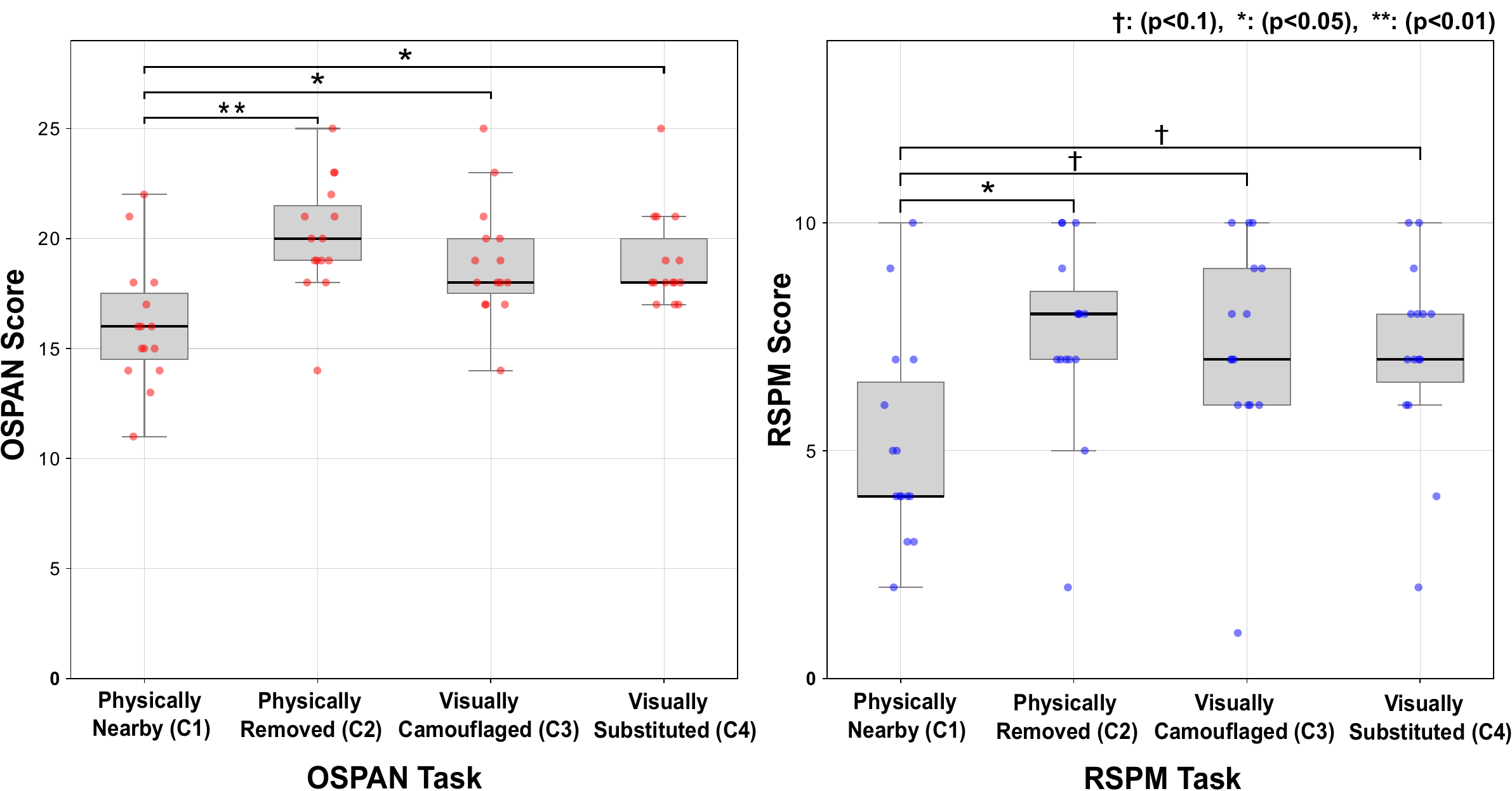}
    \caption{Scores for two cognitive capacity tasks (OSPAN and RSPM) across four conditions. The scores for C2, C3, and C4 are significantly different from C1. Additionally, the scores for C3 and C4 do not significantly differ from C2, indicating that visually canceling the phone via AR achieves similar cognitive benefits as physically removing the phone. † is marginally significant $(p<0.1)$, * is significant $(p<0.05)$, and ** is highly significant $(p<0.01)$.}
    \Description{Two side-by-side scatter plot graphs depict average OSPAN and RSPM cognitive task scores across four conditions. The OSPAN graph on the left shows higher scores for conditions C2, C3, and C4 compared to C1, with significant markers indicating statistical relevance. The RSPM graph on the right follows a similar pattern, with C2, C3, and C4 having better performance than C1.}
    \label{fig:cogni}
\end{figure*}

\subsection{Task Performance Analysis}
We conducted several analyses to evaluate the effects of different phone conditions on available cognitive capacity (see Fig. \ref{fig:cogni}). Two domain-general cognitive function metrics were used: the OSPAN and RSPM scores. These metrics were chosen for their reliance on limited-capacity attentional resources, thus serving as robust indicators for fluctuations in cognitive capacity \cite{ward2017brain}.

To assess the impact of different phone conditions on a combination of the OSPAN and RSPM scores, we used a multivariate analysis of variance (MANOVA). The Pillai's Trace statistic revealed a significant effect of different conditions on cognitive capacity ($F(6, 112) = 4.1948, p = .0008$). Subsequent univariate ANOVAs were conducted for each dependent variable. For the OSPAN task, the ANOVA revealed a significant main effect of the conditions ($F(3, 56) = 6.548, p = .0007$, $\eta^{2} = 0.259$). Similarly, the conditions had a significant main effect on RSPM scores ($F(3, 56) = 3.868, p = .0138$, $\eta^{2} = 0.172$). Bonferroni post-hoc tests were further conducted to investigate pairwise differences among the four conditions: Physically Nearby (C1), Physically Removed (C2), Visually Camouflaged (C3), and Visually Substituted (C4). C1 was set as the baseline. For the OSPAN task, significant differences were observed (Fig. {\ref{fig:cogni}}). Specifically, comparing C1 to C2, C1 to C3, and C1 to C4 showed significant differences, with $p$-values of 0.0022, 0.0415, and 0.0188, respectively. The RSPM task reflected a similar trend, showing a significance between C1 and C2 ($p = 0.0259$) and marginally significant when comparing C1 with C3 ($p = 0.0688$) and C4 ($p = 0.0909$). No significant differences were observed when comparing C2 with C3 and C4 for either task.

For the GNG task, we analyzed the effects of smartphone salience on two behavioral measures of sustained attention: omission errors and reaction time. Since the GNG data did not pass the normality test, we used the non-parametric Kruskal-Wallis test, which yielded no statistically significant effects of different conditions on either of these measures.
\subsection{Post-Interview Analysis}
This section offers insights into the visual saliency of smartphones and holograms, shifts in attention, and perspectives on future adoption and alternative designs (see Fig. \ref{fig:qualanal}).

\subsubsection{Visual Saliency of Smartphones} 
\begin{figure*}[ht]
    \centering
    \includegraphics[width=0.85\linewidth]{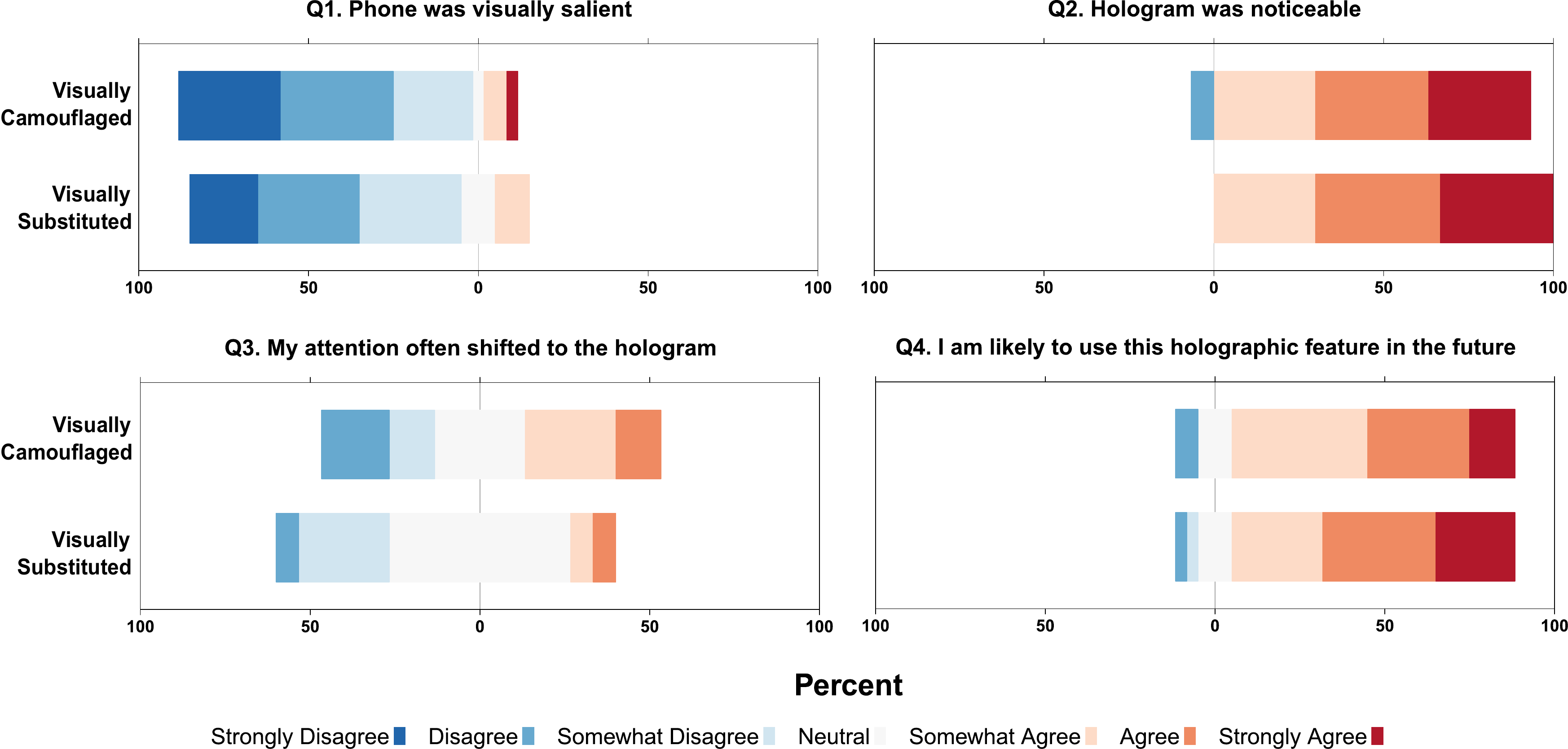}
    \caption{Participant responses on a 7-point Likert scale, comparing the two visual interventions. Overall, participants found the phone less visible when the hologram was present. The hologram was noticeable for both conditions, but the frequency of attention shifts was fairly even. The likelihood of using the holographic feature is favorable.}
    \Description{Stacked bar charts display participant responses on a 7-point Likert scale for four questions assessing the impact of Visually Camouflaged and Visually Substituted conditions on phone visibility, hologram noticeability, attention shifts, and future use likelihood. Responses suggest that the holograms reduce the phone's visual salience in both conditions. While the hologram is noticeable, it does not consistently distract participants, who indicate a favorable view towards using the holographic features in the future, signifying acceptance and perceived utility.}
    \label{fig:qualanal}
\end{figure*}

The average and standard deviation of participant responses on a 7-point Likert scale for smartphone saliency are as follows: visually camouflaged condition ($\mu = 2.37$, $\sigma = 1.40$), visually substituted condition ($\mu = 2.60$, $\sigma = 1.20$), and combined results across both conditions ($\mu = 2.48$, $\sigma = 1.31$). The distribution of responses was skewed toward the phone being perceived as 'not salient' (Fig. \ref{fig:qualanal}). A majority of the participants reported that the phone was either completely invisible or nearly so when the hologram was in place. P12 and P16 mentioned "\textit{Could not see the phone}" and "\textit{Phone was not visible at all}, respectively." P14 and P24 also indicated that the phone was visible only under certain conditions, such as "\textit{when moving their head}" or "\textit{focusing really closely at it}," respectively.

\subsubsection{Noticeability of Holograms} The average and standard deviation of participant responses on a 7-point Likert scale for the noticeability of the hologram are as follows: visually camouflaged condition ($\mu = 5.73$, $\sigma = 1.26$), visually substituted condition ($\mu = 6.03$, $\sigma = 0.80$), and combined results across both conditions ($\mu = 5.88$, $\sigma = 1.07$). The central sentiment was that the hologram was evident, often described as "\textit{bright}" or possessing a "\textit{distinct glow}." This characteristic allowed it to "stand out" even when participants were focused on other tasks, with phrases like "\textit{always in my field of view}" being recurrent. P21 also mentioned how they "\textit{could constantly see it, but not too significant later on}," indicating a habituation effect, wherein the initial allure of the hologram wore off over time. Similarly, factors like the viewing angle and lighting were mentioned by P9 and P51 as elements that could modulate the hologram's visibility.

\subsubsection{Attention Shifts to Holograms} The average and standard deviation of participant responses on a 7-point Likert scale for attention shifts are as follows: visually camouflaged condition ($\mu = 4.00$, $\sigma = 1.32$), visually substituted condition ($\mu = 3.80$, $\sigma = 0.91$), and combined results across both conditions ($\mu = 3.90$, $\sigma = 1.14$). The remarks highlighted a nuanced interplay between the hologram's visual salience and the participant's engagement with their primary tasks. P37 noted, "\textit{Curious to see if the hologram will interact during the study that may lead to distraction}," highlighting potential curiosity-induced attention shifts. On the other hand, P3 stated, "\textit{I wouldn't be affected by it, but it seems like it can form some kind of an atmosphere that may help focus}," indicating the hologram's potential to create a conducive task environment. Several participants (P18, P24, P30) referred to the brightness of the hologram, with one saying, "\textit{It was always noticeable but not distracting}." The tendency for the hologram's influence to wane over time was also noted, with P37 remarking, "\textit{At the start, it was hard to ignore, but it became more mundane as time went on}." Despite the hologram's visibility, P38 observed, "\textit{My focus remained steadfast, with only rare diversions towards the hologram}." Overall, the feedback revealed that although the hologram was often perceptually noticeable, its impact on the primary task varied and often diminished over time.

\subsubsection{Future Adoption of Holograms} The average and standard deviation of participant responses on a 7-point Likert scale for future adoption are as follows: visually camouflaged condition ($\mu = 5.27$, $\sigma = 1.21$), visually substituted condition ($\mu = 5.53$, $\sigma = 1.23$), and combined results across both conditions ($\mu = 5.40$, $\sigma = 1.23$). One of the primary motivators for using the feature is its ability to reduce distractions, particularly from smartphones. Respondents (P14, P26, P27) mention how the hologram helped them focus on tasks by obscuring their phones. Several participants (P8, P13, P24) noted that studying and working are primary use cases. For instance, P7 said, "\textit{For studying as it helps me decrease distraction when the hologram fully covers my phone}," indicating an interest in leveraging the intervention to reduce the interference of smartphones during study sessions. The use of holographic features in a work environment was also highlighted. Comments (P9, P17) like "\textit{for work to improve efficiency}" and "\textit{when I am writing my research articles because I get distracted from my phone or other objects}" suggest that professionals could use the intervention to enhance focus. The technology is seen as beneficial in a co-working setting for maintaining individual concentration, as P44 mentioned using it "\textit{in a shared workspace, to maintain a sense of individual space and concentration}." The holographic technology was also viewed as potentially helpful in creating a private ambiance in public transport and open spaces. P39 also mentioned, "\textit{In public transport, to create a more private ambiance by obscuring the crowd}," suggesting a desire for increased personal space. Interestingly, P19 noted that using AR to visually hide the phone "\textit{helped me not to be tempted to look or interact with it while the anxiety of not having it around was at a minimum}." This indicates that our interventions not only offer a way to manage distractions but also alleviate the anxiety linked to smartphone separations \cite{cheever2014out, hartanto2016smartphone, schwaiger2022impact, niu2022can, terzimehic2023tale}.

\subsubsection{Alternative Designs of Holograms} 
The feedback on alternative objects for holograms suggests a range of preferences based on functionality, aesthetic value, emotional comfort, and individual circumstances. For example, P18 noted the utility of task management by suggesting a "\textit{separate display surface to view quick-access information like appointments or to-do lists}." This statement illustrates the need for functional objects that serve as extensions of the user's productivity ecosystem. Additionally, addressing unique scenarios or needs, P55 recommended a "\textit{clear holographic calendar with all my exam dates and deadlines}," which not only adds functional value but also addresses a very particular need, thereby suggesting that specialized holographic features could benefit users in unique situations. Emotional and psychological comfort also played a significant role. P7 said they would prefer "\textit{cute pets like cats}" because they are "\textit{allergic to cat's hair, but if it would be an AR cat that could interact with me, I would be able to feel more relaxed}." This statement showcases how holograms could provide emotional sustenance, filling in gaps where real-world objects or conditions may be insufficient or harmful. P37 proposed an "\textit{animated aquarium with fish swimming around}," which they believe would be calming. This indicates that aesthetic pleasure can be beneficial even in a focused work environment.
\section{Discussion}
This study explored the cognitive effects of smartphone presence and assessed the potential of AR to mitigate these effects. Consistent with prior findings, our results confirm that the mere presence of a smartphone can hinder cognitive performance, as indicated by reduced cognitive task performance when the smartphone was physically nearby compared to when it was removed \cite{ward2017brain}. More critically, our results demonstrate that holographic AR interventions (visual camouflage and visual substitution) can counter these cognitive drawbacks. By visually canceling the smartphone, akin to auditory noise cancellation but for visual distractions, our interventions resulted in cognitive enhancement, similar to the benefits of physically removing the phone, as evidenced by the cognitive task performance results. Specifically, our study results reveal that smartphone presence impairs cognitive capacity, as reflected in OSPAN and RSPM scores, but does not affect sustained attention, measured by omission errors and reaction time in the GNG task. This result aligns with prior work, showing that even when individuals successfully maintain sustained attention by resisting the urge to check their phones, the mere presence of these devices can still diminish cognitive capacity \cite{ward2017brain}.

\subsection{Effectiveness of AR as a Visual (Noise) Cancellation Device}
The success of AR interventions lies in their ability to reduce the salience of the smartphone, which would otherwise compete for attentional resources. Smartphones are high-priority stimuli due to their personal relevance and chronic salience, akin to hearing one's name or a baby's cry, which automatically capture attention \cite{shiffrin1977controlled, roye2007personal}. The visual cancellation provided by AR effectively diminishes this salience, reducing the "gravitational pull" the smartphone exerts on attention and freeing cognitive resources for task-related processes. This aligns with theories of limited-capacity attentional systems, which emphasize that occupying cognitive resources to inhibit irrelevant but salient stimuli impairs performance on other tasks \cite{kahneman1973attention, lavie2004load}. 

Moreover, participants reported a habituation effect to the holograms, indicating that while the overlays were initially noticeable, they did not persistently draw attention. Qualitative feedback suggests that participants did not "forget" the smartphone but rather perceived it as "completely invisible" or "non-salient" during the AR interventions. This reflects the power of the interventions to modulate the environment in ways that reduce the smartphone's cognitive impact. Unlike conscious suppression of distraction, which requires active cognitive effort \cite{parr2023cognitive}, the holographic overlays likely offloaded the cognitive burden associated with suppressing attention to the smartphone. This contrasts with static interventions, such as covering the phone with a piece of paper, which may remain conspicuous due to their incongruence with the surrounding environment. 

AR dynamically adapts to the context, visually blending with the environment in ways static solutions cannot achieve. Our study highlights two complementary mechanisms that drive the cognitive benefits of AR-based distraction mitigation: salience reduction and attentional guidance. Salience reduction occurs when AR interventions diminish the perceptual distinctiveness of distracting objects, such as smartphones, thereby reducing their ability to automatically capture attention. This mechanism is most evident in the visual camouflage condition, where the phone is blended into the background, eliminating visual competition. Attentional guidance, on the other hand, is achieved through visual substitution, where holograms (e.g., a holographic book) not only obscure the distracting object but also act as contextual cues to focus users on task-relevant information. Unlike passive solutions like covering the phone with an object, AR dynamically blends with the environment and maintains perceptual consistency.

These dual mechanisms align with theories of attentional prioritization, which emphasize the interplay between stimulus salience and goal relevance in guiding attention \cite{corbetta2002control}. By reducing the salience of irrelevant stimuli (visual camouflage) and reinforcing goal-relevant anchors (visual substitution), AR achieves robust cognitive benefits. This dual approach ensures that users experience reduced attentional competition from irrelevant stimuli while also benefiting from perceptual guidance toward task-relevant information. Collectively, this synergy of salience reduction and attentional guidance surpasses the capabilities of simpler static interventions, offering a comprehensive strategy for distraction mitigation.

\subsection{Practical Implications \& Potential Use Case}
It is important to clarify that our solution is not limited to smartphones. We chose smartphones as an example of a common distraction, as they are omnipresent. However, our proposed method can be generalized to various situations. For instance, it can help improve focus and concentration by visually diminishing clutter in a room or minimize distractions in an open office environment by obscuring irrelevant visual stimuli. This adaptability is a crucial aspect of our solution. While a smartphone can be physically removed by its owner, physically removing other objects or people requires consent and effort, which may not always be possible. Additionally, physically removing smartphones from immediate reach isn't always viable or desirable due to the fear of missing out, which escalates with prolonged separation. Our intervention surpasses these limitations as it only visually diminishes distractions from the user's FOV.

One compelling application can be related to addressing prevalent issues such as glossophobia or stage fright \cite{khurpade2020effect}. The first step of the application involves techniques to visually cancel out elements within the audience that may cause stress or anxiety for the speaker. This could include faces exhibiting judgmental expressions or other distracting visual cues. After establishing this holographic visual dominance, the reduced or minimized elements can be replaced with more comforting visuals using visual substitution. For example, speakers equipped with the AR HMD could then see these visually diminished areas transformed into faces of friends or family \cite{arias2021controlling}, manipulated with filters \cite{leong2023picture}, or replaced with calming landscapes \cite{paredes2018driving}. Alternatively, the audience could be transformed into non-judgmental figures, further diminishing the stress associated with the perception of judgment \cite{zajonc1965social}. This combined approach may be promising for refining visual effects and could set a new direction for developing holographic display technology.

Another promising direction involves introducing user-customizable experiences, a feature whose importance has been well-established in prior research \cite{kim2022shared,wunderlich2013high,marathe2011drives}. This would allow users to personally select their desired overlays for distracting objects. For example, P7's preference for a hologram of "cute pets like cats" due to an allergy to cat hair exemplifies how such customization could fulfill emotional needs in cases where real-world options are impractical. Future work could explore using holograms as emotional support tools, especially where real-life alternatives are impractical or harmful. This approach would blend visual camouflage and substitution techniques not just to reduce distractions but also to enhance user well-being. The initial stage would involve visual camouflage to effectively reduce the smartphone's visibility. After establishing this visual dominance, a context-appropriate object could be overlaid using visual substitution. This personalization could be enhanced by machine learning, predicting and suggesting holographic overlays based on users' historical behavior and preferences \cite{carrozzi2019s}. Such predictive analytics aims to generate holographic overlays that resonate with the user's inclinations and induce states of cognitive focus or relaxation as required. By adapting to each user's unique preferences, we believe this personalized approach would elevate the practicality of our system.

\subsection{Applicability to Spatial AR and Societal XR \& Ethical Considerations}
Our proposed AR interventions, designed to visually diminish distractions, offer potential for integration into extended reality (XR) paradigms that do not require HMDs. Specifically, the principles behind our methods can extend to Spatial Augmented Reality (SAR) and Societal XR frameworks \cite{bimber2005spatial, gorlich2022societal}. SAR utilizes environmental technologies like projectors to overlay virtual elements onto physical spaces, eliminating the need for users to wear technology \cite{lee2019applying, jung2019augmented}. Our approach aligns with SAR by projecting contextually relevant imagery or occlusions to visually conceal distracting objects. While SAR systems hold the potential for concealing flat objects like smartphones, projecting onto larger or more three-dimensional objects may present challenges. This is due to the spatial relationship between the object, the projector, and the user’s viewpoint, which can result in portions of the object remaining visible if not fully within the projector's line of sight. Nevertheless, for flat objects and controlled environments, SAR remains a promising avenue for distraction mitigation. Similarly, Societal XR envisions XR technologies embedded in public and shared spaces, promoting accessibility for a broader audience without relying on wearable technology. Our techniques could contribute to this paradigm by minimizing distractions in communal spaces such as open offices, libraries, and classrooms. These systems can benefit from our findings, as projection-based AR in physical spaces requires thoughtful design considerations to effectively reduce visual distractions. Overall, our work supports efforts to mitigate distractions without XR glasses, opening new possibilities for AR applications in shared and public environments.

However, two concerns arise when implementing such interventions: reduced user awareness and potential accidents \cite{kim2024collision}. For example, diminishing the visibility of objects could unintentionally obscure safety-relevant cues, such as emergency notifications or hazardous items. To address this, AR systems should incorporate context-aware adaptations that dynamically prioritize the visibility of essential information, ensuring users maintain situational awareness even as distractions are minimized. Similarly, in environments like workplaces or classrooms, overly aggressive distraction suppression could reduce awareness of interpersonal interactions or environmental changes. By integrating adaptive features that balance distraction minimization with context sensitivity, our interventions can ensure that users remain connected to their surroundings while benefiting from enhanced focus.

As XR technologies evolve, ethical concerns about their societal impact must remain at the forefront. A concern is that by selectively manipulating environmental elements, whether objects or people, we risk altering the human experience and undermining ethical norms in social interaction. This selective manipulation affects collective social realities, raising ethical questions about perception, who decides what is diminished, and the broader societal impacts of these choices. Adding a cultural perspective, the potential for misuse is highlighted in popular media. For instance, in the Black Mirror episode “White Christmas” (S2E4), individuals are "blocked" and rendered invisible to society through mediated reality contact lenses \cite{muller2019we}. When blocked, the person appears as a distorted image, ostracizing them from any social interaction. Such a dystopian vision raises ethical concerns about consent, privacy, and the consequences of selectively excluding people from shared reality. As this technology matures, a thorough ethical examination is imperative to address the concerns arising from modifying shared social environments.

\section{Limitations \& Future Work}
Despite promising findings from our AR interventions, there exist limitations. First, the HL2 offers a limited FOV, prompting us to limit users' positions during the experiment. This technical constraint of the device is notable when considering the broader implications for cognitive performance \cite{caluya2022does}. Future studies should explore AR devices with wider FOVs to better understand their impact on cognitive performance. We also envision that this limitation will subside as more advanced and ergonomic AR devices emerge \cite{gopakumar2024full}. The second limitation is the limited occlusion capability of AR devices. Unlike VR displays, the virtual elements (e.g., holograms) often appear translucent due to AR's additive display. Hence, AR perception studies are typically conducted within rigorously controlled environments to maximize the contrast and brightness between virtual and real elements. Similarly, we addressed this by dimming the ambient light to enhance the contrast between the virtual and real worlds, but this method has limitations. Generally, modifying the lighting conditions is an impractical approach for real-world applications where lighting conditions can vary significantly. The lack of effective occlusion could compromise the full potential of the visual interventions. However, we reemphasize that our decision to use AR instead of VR was intentional, as AR allows real-time interaction with the real world without distortion (Sec. \ref{sec:ar}). While VR is valuable for fully immersive experiences, its passthrough systems are not feasible for detailed cross-reality work, creating a sense of detachment from real-world surroundings and reactions \cite{wentzel2024switchspace}. Moreover, achieving full DR solutions in 3D space in real-time with varying objects remains a significant challenge, even with VR devices. We anticipate that advancements in the occlusion capabilities of head-mounted AR devices will provide valuable insights for achieving a more immersive user experience \cite{kiyokawa2001optical, avveduto2017real, mori2018brightview}.

Additionally, since participants had different phones and settings, it was impractical to control for various disruptions and notifications (e.g., mute, vibration, sound, brightness, flashing), which could introduce further confounds. To address this, we instructed participants to turn off their phones, ensuring consistency across conditions. This decision was based on prior research demonstrating that cognitive performance remains unaffected by the phone's power condition \cite{ward2017brain, skowronek2023mere}. However, we recognize that this decision limits the ecological validity of the study, as in real-world scenarios, phones are typically active, and notifications can influence attention. In future work, experiments could include conditions where phones remain turned on but are placed on silent mode, reflecting common real-world scenarios where users typically do not turn off their phones completely \cite{saling2016you}. Such an approach could help investigate whether AR-based interventions reduce the cognitive pull of an active phone, even when participants know they could still check it.

We hypothesize that DR interventions would be even more effective when phones remain on, as they would reduce the visual salience of active notifications, thereby mitigating their distracting influence. In contrast, the baseline condition where the phone is physically present without AR intervention lacks any mechanism to counteract the visual prominence of notifications. While positioning phones face down could serve as an alternative to obscure notifications, the phone itself remains visible, and its presence may still be tempting. Our DR interventions offer a potential advantage by transforming the phone’s physical features into an alternative holographic representation, which could further diminish its perceptual presence in ways that baselines cannot achieve. Future AR interventions could also incorporate adaptive features that dynamically respond to changes in the phone’s visual state, such as illuminated screens or flashing alerts. For instance, holographic overlays could intensify or reconfigure in real time to obscure the increased brightness of notifications, ensuring consistent distraction reduction. Exploring these dynamics in future research would provide valuable insights into how DR systems can maintain robustness and adaptability in real-world scenarios where phones are actively used.

\section{Conclusion}
Smartphones, now an indispensable part of our daily lives, bring along cognitive drawbacks simply by being present. The \textit{brain drain} phenomenon reports that having one's smartphone within sight can drain cognitive resources, compromising task performance. While effective, removing smartphones from immediate reach isn't always practical or desirable, especially with the fear of missing out increasing over prolonged separation. This challenge prompts us to question whether AR could reduce their cognitive distractions by visually canceling smartphones. We used the Microsoft HoloLens 2 to visually camouflage or substitute the phone via AR to address this. Our results showed that both interventions improved cognitive performance to levels similar to physically removing the phone. Although our study used smartphones, the approach is generalizable for visually canceling out any objects that may be distracting. This introduces new design perspectives, showing AR's potential not only to augment but also to reduce distractions, with practical implications for enhancing cognitive environments and managing daily distractions.

\begin{acks}
We thank the participants for their time and participation. This work is supported by the Canada Foundation for Innovation (CFI) John R. Evans Leaders Fund (JELF) grant no. 44170, and Natural Sciences and Engineering Research Council of Canada (NSERC) Discovery Grant no. RGPIN-2023-04148.
\end{acks}
\bibliographystyle{ACM-Reference-Format}
\bibliography{0_export}
\end{document}